\documentclass{article}
\usepackage{spconf}

\def\BibTeX{{\rm B\kern-.05em{\sc i\kern-.025em b}\kern-.08em
    T\kern-.1667em\lower.7ex\hbox{E}\kern-.125emX}}

\include{epsf}
\include{epsfrot}
\usepackage{times}
\usepackage{latexsym}
\usepackage{amsmath}
\usepackage{amsfonts}
\usepackage{pstricks}
\usepackage{amssymb}
\usepackage{amsxtra}
\usepackage{pstricks}
\usepackage{epsfig}
\usepackage{color}
\usepackage{graphics}
\usepackage{psfrag}

\title{Energy-Efficient Joint Estimation in Sensor Networks:\\ Analog vs. Digital}

\name{Shuguang Cui$^1$,\ \ Jin-Jun Xiao$^2$,\ \ Andrea J.
Goldsmith$^1$,\ \ Zhi-Quan Luo$^2$,\ \ and \ \ H. Vincent Poor$^3$
\thanks{This research is supported in part by funds from National Semiconductor and Toyota
Corporation, by the Natural Sciences and Engineering Research
Council of Canada, Grant No.\ OPG0090391, 
by the National Science Foundation, Grant No.\
DMS-0312416, by the Office of Naval Research under Grant
N00014-03-1-0102.}
\address{\small $^1$Wireless System
Lab, Department of Electrical Engineering, Stanford University. 
\\ \small $^2$Department of Electrical and Computer Engineering, University of Minnesota. 
\\ \small $^3$Department of Electrical Engineering, Princeton University. \vspace{-20pt}
}\vspace{-10pt} }

\newcommand{\ie}{i.e.}

\newcommand{\nr}{\nonumber}

\begin{document}
\ninept

\maketitle \thispagestyle{empty} \pagestyle{empty}

\begin{abstract}

Sensor networks in which energy is a limited resource so that
energy consumption must be minimized for the intended application
are considered.  In this context, an energy-efficient method for
the joint estimation of an unknown analog source under a given
distortion constraint is proposed. The approach is purely analog,
in which each sensor simply amplifies and forwards the
noise-corrupted analog observation to the fusion center for joint
estimation. The total transmission power across all the sensor
nodes is minimized while satisfying a distortion requirement on
the joint estimate. The energy efficiency of this analog approach
is compared with previously proposed digital approaches with and
without coding. It is shown in our simulation that the analog
approach is more energy-efficient than the digital system without
coding, and in some cases outperforms the digital system with
optimal coding.
\end{abstract}


\section{Introduction}

\begin{figure}[!h]
\centering
      \scalebox{0.25}{\includegraphics{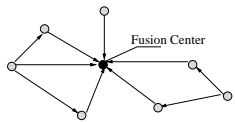}}
       \vspace{-10pt} \caption{Sensor network with a fusion center.}
    \label{wsn_graph} \vspace{-10pt}
\end{figure}

A typical Wireless Sensor Network (WSN), as shown in
Fig.~\ref{wsn_graph}, consists of a fusion center and a large
number of geographically distributed sensors. The sensors
typically have limited energy resources and communication
capability. Each sensor in the network makes an observation of the
target of interest, generates a local signal (either analog or
digital), and then sends it to the fusion center where the
received sensor signals are combined to produce a final estimate
of the observed signal. Sensor networks of this type are
well-suited for situation awareness applications such as
environmental monitoring and smart factory instrumentation.

\vspace{-5pt}

Decentralized estimation has been studied first in the context of
distributed control~\cite{castanon}, distributed
tracking~\cite{willsky}, and most recently in wireless sensor
networks~\cite{papadopoulos}. Among these studies, it is usually
assumed that the joint distribution of the sensor observations is
known. In practical systems, the probability density function
(pdf) of the observation noise is hard to characterize, especially
for a large scale sensor network. This motivates us to devise
signal processing algorithms that do not require the knowledge of
the sensor noise pdf. Recently, universal decentralized estimation
schemes (DES) without the knowledge of noise distribution have
been proposed in~\cite{luo1} and~\cite{luo3}. In~\cite{luo1}, the
author considered the universal DES in a homogeneous sensor
network where sensors have observations of the same quality, while
in~\cite{luo3}, the universal DES in an inhomogeneous sensing
environment was considered. These proposed DESs require each
sensor to send to the fusion center a short discrete message with
length decided by the local Signal to Noise Ratio (SNR), while the
performance is guaranteed to be within a constant factor of that
achieved by the Best Linear Unbiased Estimator (BLUE). An
assumption in these proposed schemes is that the channels between
sensors and the fusion center are perfect, and all messages are
received by the fusion center without any distortion. However, due
to power limitations and channel noise, the signal sent by each
individual sensor to the fusion center will be corrupted.
Therefore, the goal of the transmission system design for the
joint estimation problem is to minimize the effect of the channel
corruption while consuming the minimum amount of power at each
node.

Historically, if the sensor observation is in analog form, we have
two main options to transmit the observation from the sensors to
the fusion center: analog or digital communication. For the analog
approach, we keep the observation signal analog and further use
analog modulation schemes to transmit the signal, which is also
called an amplify-and-forward approach. In the digital approach,
we digitize the observation into bits, possibly apply channel
coding, then use digital modulation schemes to transmit the data.
It is well known (\cite{Goblick},~\cite{Gastpar}) that for a
single Gaussian source with an AWGN channel, the
amplify-and-forward approach is optimal. However, for an arbitrary
source with multiple observations and multiple transmission
channels, it is unclear which approach will lead to a smaller
power consumption while meeting the distortion requirements at the
fusion center.

Another severe challenge facing sensor networks is the hard energy
constraint. Since each sensor is equipped with only a small-size
battery, for which replacement is very expensive if not impossible,
the power consumption must be minimized to increase the network
lifetime~\cite{Shuguang3}. Therefore, energy efficiency can be used
as a performance criterion to evaluate different sensor network
designs under the same distortion requirement.
In~\cite{Jinjun_Shuguang1}, a digital system is proposed to minimize
the total power consumption for the joint estimation problem, where
each sensor quantizes the analog observation into digital bits and
transmits the bits to the fusion center with uncoded MQAM. The
number of quantization bits for each sensor is optimized with the
target of minimizing the total transmission power across all the
sensor nodes to achieve a given distortion. In this paper, we
propose an analog counterpart for the same system and compare the
energy efficiency between the analog approach and the digital one.
As in~\cite{Jinjun_Shuguang1}, we assume that the observed signal is
analog and bounded, the fusion center deploys the best unbiased
linear estimator, the observation noise is uncorrelated across
different sensors, and only the variance of the observation noise is
known.

Our paper is organized as follows. Section II discusses the
problem formulation. Section III compares the energy efficiency
between the analog approach and the digital approach via some
numerical examples. Section IV summaries our conclusions.

\vspace{-5pt}
\section{Minimum Power Analog information collection}\vspace{-5pt}

We assume that there are $K$ sensors and the observation $x_k(t)$ at
sensor $k$ is represented as a random signal $\theta(t)$ corrupted
with the observation noise $n_k(t)$: $x_k(t)=\theta(t)+n_k(t)$. Each
sensor transmits the signal $x_k(t)$ to the fusion center where
$\theta(t)$ is estimated from the $x_k(t)$'s, $k=1,\cdots,K$. We
further assume that $n_k(t)$ is of unknown statistics and the
amplitude of $x(t)$ is bounded within $[-W,W]$, which is defined by
the sensing range of each sensor. For simplicity we assume $W=1$,
but our analysis can be easily extended to any values. We also
assume that the network is synchronized, which may be enabled by
utilizing beacon signals in a separate control channel.

We assume that $x_k(t)$ is also band-limited and the information
is contained within the frequency range $[-B/2,B/2]$. We consider
an analog Single Side-Band (SSB) system with a coherent
receiver~\cite{Haykin1}. The transmitted signal is given by
\begin{equation}
y_t(t)=2\sqrt{\alpha}\cos{(\omega_ct)}x(t)+2\sqrt{\alpha}\sin{(\omega_ct)}\hat{x}(t),
\end{equation}
for which the average transmission power is
\begin{equation}\label{Eq_power1}
P=4\alpha{P_x}\le4\alpha{W^2}
\end{equation}
where $4\alpha$ is the transmitter power gain and $P_x$ is the
peak power of $x(t)$.

\begin{figure}[!h]
\psfrag{theta}{$\theta$} \psfrag{b1}{$2\sqrt{\alpha_1}$}
\psfrag{b2}{$2\sqrt{\alpha_2}$}  \psfrag{bK}{$2\sqrt{\alpha_K}$}
\psfrag{a1}{$\sqrt{g_1}$} \psfrag{a2}{$\sqrt{g_2}$}
\psfrag{aK}{$\sqrt{g_K}$} \psfrag{n1}{$n_1$} \psfrag{n2}{$n_2$}
\psfrag{n3}{$n_K$} \psfrag{nc1}{$n_{c1}$} \psfrag{nc2}{$n_{c2}$}
\psfrag{ncK}{$n_{cK}$} \psfrag{yk}{$y_1, \cdots, y_K$}
\begin{center}
\hspace{-45pt}
 {\includegraphics[width=2.8in, height=1.0in]{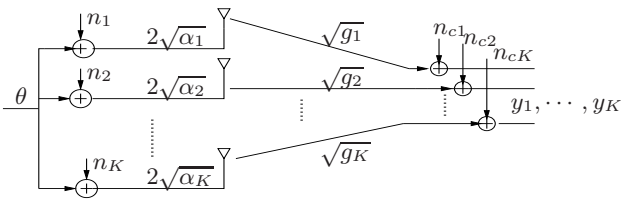}}
\end{center}
\vspace{-15pt} \caption{Amplify and Forward}\label{Fig_sensor_fa}
\vspace{-15pt}
\end{figure}

The received signal at the fusion center is given by
\begin{equation}
y_r(t)=2\sqrt{\alpha}\sqrt{g}\cos{(\omega_ct)}x(t)+2\sqrt{\alpha}\sqrt{g}\sin{(\omega_ct)}\hat{x}(t)
+{n}_c(t)
\end{equation}
where $g$ is the channel power gain, $\hat{x}(t)$ is the Hilbert
transform of $x(t)$, and ${n}_c(t)$ is the channel AWGN. Hence, at
the output of the coherent detector the signal is
\begin{equation}
y(t)=\sqrt{\alpha}\sqrt{g}x(t)+\frac{1}{2}n_{c}^I(t)\cos(\pi{\frac{B}{2}t})
+\frac{1}{2}n_{c}^Q(t)\sin(\pi{\frac{B}{2}t}),
\end{equation}
where $n_{c}^I(t)+jn_{c}^Q(t)$ is the complex envelope of
$n_c(t)$. After passing $y(t)$ through a low-pass filter and
sampling the baseband signal at a sampling rate $B$, we can obtain
an equivalent discrete-time system. Since we have $K$ such
sensors, the overall system is shown in Fig.~\ref{Fig_sensor_fa}.
The $K$ transmitters share the channel via Frequency Division
Multiple Access (FDMA), which has the same spectral efficiency as
the Time Division Multiple Access (TDMA) that is used
in~\cite{Jinjun_Shuguang1}.

The received signal vector at each time instance is given by
\begin{equation}
\mathbf{y}=\mathbf{h}\theta+\mathbf{v},
\end{equation}
where \vspace{-10pt}\begin{eqnarray} \mathbf{y}&=&\left[y_1,
y_2,\cdots,y_K\right]^{\dag}, \nonumber \\
\mathbf{h}&=&\left[{\sqrt{\alpha_1g_1}}, {\sqrt{\alpha_2g_1}},
\cdots, {\sqrt{\alpha_Kg_K}}\right]^{\dag}, \nonumber
\\ \mathbf{v}&=& \left[{\sqrt{\alpha_1g_1}}n_1+n_{c1},
\cdots,{\sqrt{\alpha_Kg_K}}n_k+n_{cK}\right]^{\dag}, \nonumber
\end{eqnarray}
and $\dag$ means transpose.

According to~\cite{Mendel1}, the best linear unbiased estimator
(BLUE) for $\theta$ is given by
\begin{eqnarray}
\hat{\theta}&=&[\mathbf{h}^{\dag}\mathbf{R}^{-1}\mathbf{h}]^{-1}\mathbf{h}^{\dag}\mathbf{R}^{-1}\mathbf{y}
\nonumber \\
&=&\left(\sum_{k=1}^{K}\frac{{\alpha_kg_k}}{\sigma_k^2{\alpha_kg_k}+\xi_{k}^2}\right)^{-1}
\sum_{k=1}^{K}\frac{{\sqrt{\alpha_kg_k}}y_k}{\sigma_k^2{\alpha_kg_k}+\xi_{k}^2},
\end{eqnarray}
where the noise variance matrix $\mathbf{R}$ is a diagonal matrix
with $R_{kk}={\sigma_k^2{\alpha_k}{g_k}+\xi_{k}^2}$ with
$\sigma_k^2$ the variance of the sensor observation noise
$n_k(t)$, $k=1,\cdots,K$. The channel noise variance $\xi_k^2$ is
defined by the noise power spectral density and the bandwidth $B$.


The mean squared error of this estimator is given
as~\cite{Mendel1} \vspace{-5pt}\begin{eqnarray}
Var[\hat{\theta}]&=&[\mathbf{h}^{\dag}\mathbf{R}^{-1}\mathbf{h}]^{-1}
\nonumber \\ &=&
\left(\sum_{k=1}^{K}\frac{{\alpha_kg_k}}{\sigma_k^2{\alpha_kg_k}+\xi_{k}^2}\right)^{-1}.
\end{eqnarray}

According to Eq.~(\ref{Eq_power1}), the transmit power for node
$k$ is bounded by $4W^2\alpha_k$. Therefore, the minimum power
analog information collection problem can be cast as \vspace{-8pt}
\begin{eqnarray} \min && \sum_{k=1}^{K}W^2\alpha_k
\nr
\\ \mbox{s.~t.} &&
\left(\sum_{k=1}^{K}\frac{{\alpha_kg_k}}{\sigma_k^2{\alpha_kg_k}+\xi_{k}^2}\right)^{-1}\le{D_0}
\nonumber
 \\ && {\alpha_{k}}\ge0,
\hspace{0.3cm}k=1,\cdots,K\nr
\end{eqnarray}
where $D_0$ is the distortion target. However, this problem is not
convex over the $\alpha_k$'s.

Let us define
$$r_k=\frac{{\alpha_kg_k}}{\sigma_k^2{\alpha_kg_k}+\xi_{k}^2}=\frac{1}
{\sigma_k^2+\frac{\xi_{k}^2}{g_k\alpha_k}}.$$ Then the above
optimization problem is equivalent to
\vspace{-8pt}\begin{eqnarray} \min && \sum_{k=1}^{K}W^2\alpha_k
\nr
\\ \mbox{s.~t.} &&
\sum_{k=1}^{K}r_k\ge\frac{1}{D_0} \nr
\\ && r_k=\frac{1}
{\sigma_k^2+\frac{\xi_{k}^2}{g_k\alpha_k}},
\hspace{0.3cm}{\alpha_{k}}\ge0, \hspace{0.3cm}\forall{k}, \nr
\end{eqnarray}
where we see that the variable $\alpha_k$ can be completely
replaced by a function of $r_k$. Therefore, the problem can be
transformed into a problem with variables $\{r_1, r_2,\dots,
r_K\}$ shown as follows:
\begin{eqnarray}\label{final_prob0}
\min &&
\sum_{k=1}^{K}\frac{W^2\xi_{k}^2}{g_k}\left(\frac{r_k}{1-r_k\sigma_k^2}\right)\nr
\\ \mbox{s.~t.} &&
\sum_{k=1}^Kr_k\ge\frac{1}{D_0};\hspace{0.3cm}
0\le{r_{k}}<{\frac{1}{\sigma_k^2}},\hspace{0.3cm}\forall{k}
\end{eqnarray}
which is convex over $r_k$. The upper limit on $r_k$ in the second
constraint is due to the fact that $r_k=1/
{\left(\sigma_k^2+\frac{\xi_{k}^2}{g_k\alpha_k}\right)}$ and
$\frac{\xi_{k}^2}{g_k\alpha_k}\ge0$.


Now we solve Eq.~(\ref{final_prob0}). Its Lagrangian $G$ is given
as
\begin{eqnarray}
G(L,\lambda_0)&=&\sum_{k=1}^{K}\frac{W^2\xi_{k}^2}{g_k}\left(\frac{r_k}{1-r_k\sigma_k^2}\right)
\nr\\&&+\lambda_0\left(\frac{1}{D_0}- \sum_{k=1}^Kr_k\right)
\end{eqnarray}
for $0\le{r_{k}}\le{\frac{1}{\sigma_K^2}}, \forall{k}$, which leads
to the following Karush-Kuhn-Tucker (KKT) conditions~\cite{Boyd1}:
\begin{eqnarray}
\frac{W^2\xi_{k}^2}{g_k}\frac{1}{(1-r_k\sigma_k^2)^2}-\lambda_0=0,
&& \forall{k}\nr \\ \sum_{k=1}^Kr_k-\frac{1}{D_0}=0 &&\nr
\end{eqnarray}
for $0\le{r_{k}}\le{\frac{1}{\sigma_K^2}}, \forall{k}$. Without
loss of generality, we rank the channel quality such that
$\frac{\xi_1^2}{g_1}\le \frac{\xi_2^2}{g_2}\le\ldots\le
\frac{\xi_K^2}{g_K}$ where we call the quantity
$\frac{g_k}{\xi_k^2}$ the channel SNR, and we define
\begin{eqnarray}\label{Eq_threshold1}
f(M)=\frac{\sqrt{\frac{\xi_{M}^2}{g_M}}A(M)} {B(M)},\quad
\mbox{for\ }1\le{M}\le{K},
\end{eqnarray}
where $A(M)=\displaystyle\sum_{m=1}^{M}
\frac{\sqrt{\frac{\xi_{m}^2}{g_m}}}{\sigma_m^2}$ and
$B(M)=\displaystyle\sum_{m=1}^{M}\frac{1}
{\sigma_m^2}-\frac{1}{D_0}$.

Let us find $K_1$ such that $f(K_1)<1$ and $f(K_1+1)\ge1$. Using
the same techniques as in~\cite{Jinjun_Shuguang1}, we can show
that this $K_1$ is unique unless $f(M)<1$ for all $1\le M\le K$,
in which case we take $K_1=K$. Then the KKT conditions give
$\lambda_0=\left(\frac{WA(K_1)}{B(K_1)}\right)^2$ and
\begin{equation}\label{Eq_result}
r_k^{opt}=\frac{1}{\sigma_k^2}\left(1-W\frac{\sqrt{\frac{\xi_{k}^2}{g_k}}}{\sqrt{\lambda_0}}\right)^+,
\hspace{0.3cm}\forall\ {k}
\end{equation}
where $(x)^+$ equals $0$ when $x<0$, and otherwise is equal to
$x$.

Hence, by definition, we have
\begin{eqnarray}\label{Eq_powerallocation1}
\alpha^{opt}_k&=&\frac{\xi_{k}^2}{g_k}\frac{r^{opt}_k}{1-\sigma_k^2r^{opt}_k}
\nonumber
\\
&=&\frac{\xi_{k}^2}{g_k\sigma_k^2}\left(\sqrt{\frac{g_k}{\xi_{k}^2}}\eta_0-1\right)
\hspace{0.3cm}k=1,\cdots,K_1,
\end{eqnarray}
and $\alpha_k=0$ otherwise, where $\eta_0=A(K_1)/B(K_1)$.

Therefore, the optimal power allocation strategy is divided into
two steps. In the first step, a threshold for $k$ is obtained
according to Eq.~(\ref{Eq_threshold1}). For channels with SNR
worse than this threshold, the corresponding sensor is shut off
and no power is wasted. For the remaining active sensors, power
should be assigned according to Eq.~(\ref{Eq_powerallocation1}).
From Eq.~(\ref{Eq_powerallocation1}) we see that when the channel
is fairly good, \ie, $\sqrt{\frac{g_k}{\xi_{k}^2}}\eta_0\gg1$, we
have
$\alpha^{opt}_k\propto{\sqrt{\frac{\xi_{k}^2}{g_k}}\frac{\eta_0}{\sigma_k^2}}$,
which means the optimal solution is inversely proportional to the
square root of the channel SNR. When
$\sqrt{\frac{g_k}{\xi_{k}^2}}\eta_0$ is close to one, the optimal
solution may no longer have such properties. For all channel
conditions, the power is scaled by the factor
$\frac{1}{\sigma_k^2}$, which means that more power is used to
transmit the signals from the sensors with better observation
quality.

\begin{figure}[!h]
\begin{center}
  {\includegraphics[width=1.6in, height=1.5in]{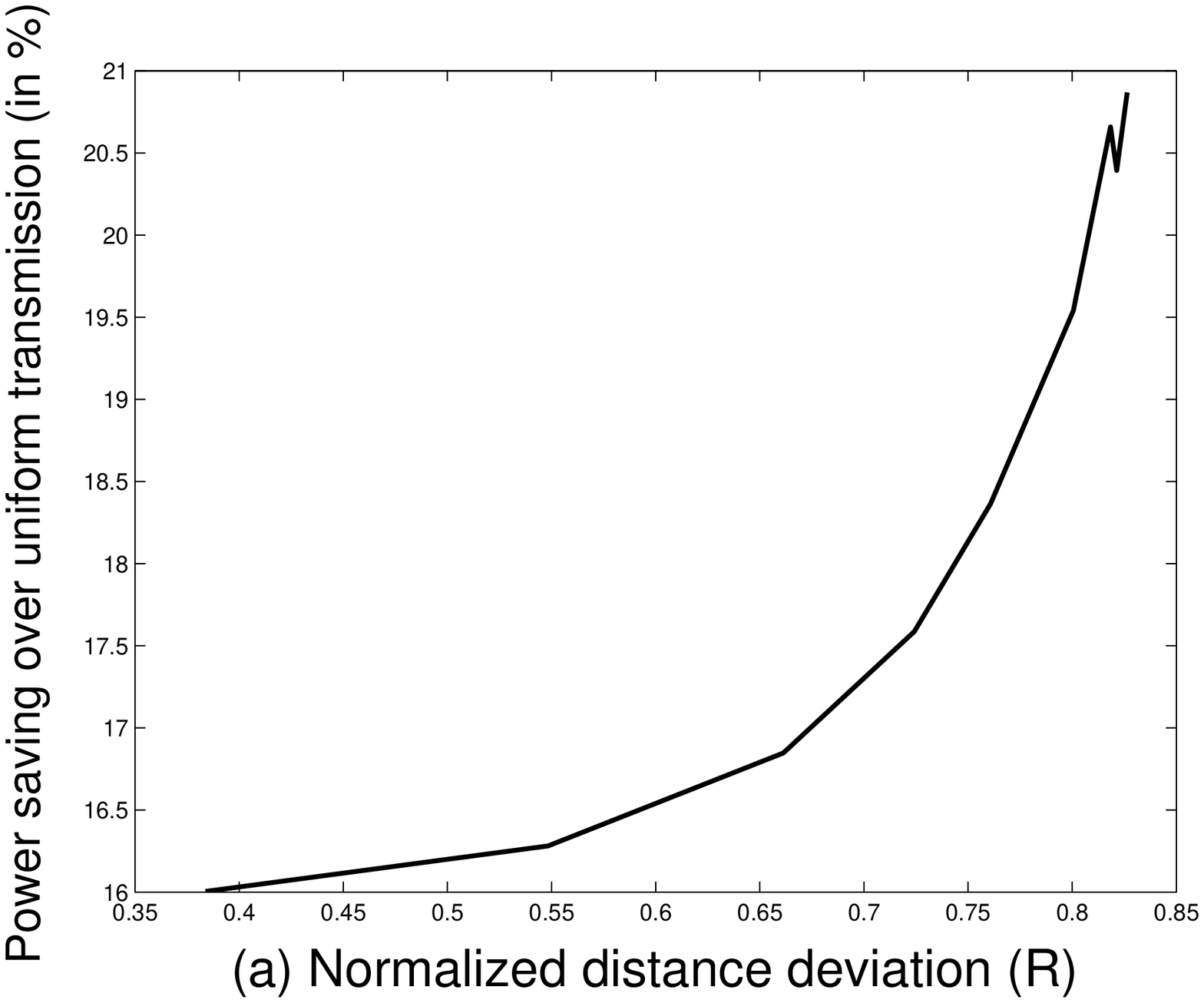}}
  {\includegraphics[width=1.6in, height=1.4in]{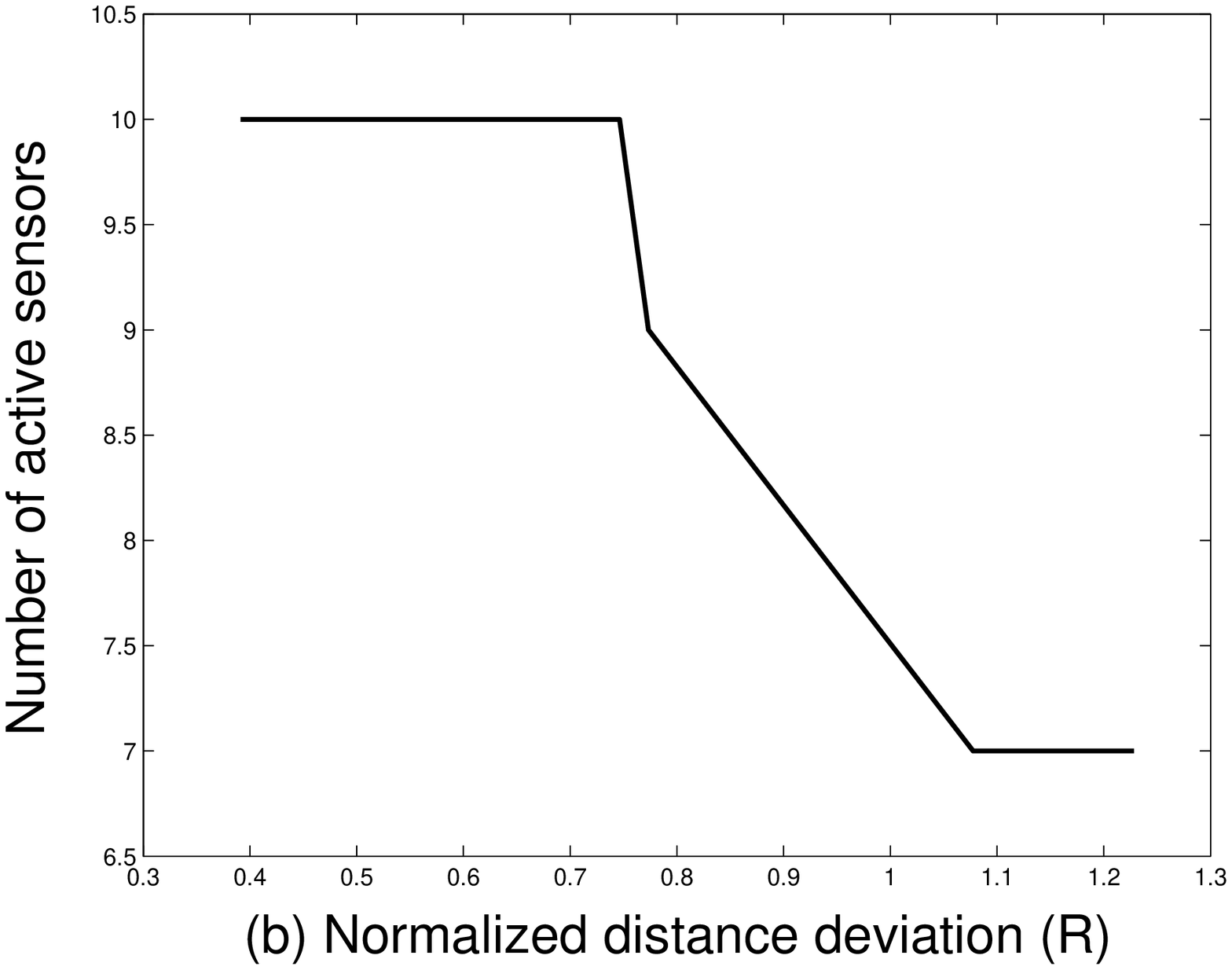}}
\end{center}
\vspace{-15pt} \caption{(a) Power savings of optimal power
allocation vs. uniform power; (b) Number of active sensors versus
distance deviation}\label{Fig_power_saving}\vspace{-5pt}
\end{figure}

We now solve the optimization problem for some specific examples.
We assume that the channel power gain $g_k=\frac{G_0}{d_k^{3.5}}$
where $d_k$ is the transmission distance from sensor $k$ to the
fusion center and $G_0=-30$~dB is the gain at $d=1$~m. As
in~\cite{Jinjun_Shuguang1}, we generate $\sigma_k^2$ uniformly
within the range $[0.01, 0.08]$. We take $B=10$~KHz and
$\xi_k^2=-90$~dBm, $k=1,\cdots,K$. For an example with $100$
sensors, Fig.~\ref{Fig_power_saving}~(a) shows the relative power
savings compared with the uniform transmission strategy where all
the sensors use the same transmission power to achieve the given
distortion target. The relative power savings is plotted as a
function of $R=\frac{\sqrt{\mathbf{Var}(d)}}{\mathbf{E}(d)}$, the
distance deviation normalized by the mean distance. For each value
of $R$, we average the relative power savings over $100$ random
runs where in each run the $d_k$'s are randomly generated
according to the given $R$. As expected, a larger variation of
distance and corresponding channel quality leads to a higher power
savings when this variation is exploited with optimal power
allocation.

In Fig.~\ref{Fig_power_saving}~(b), the number of active sensors
over $R$ is shown for an example with $10$ sensors. We see that
when the transmission distances for different sensors span a wide
range of values (\ie, $R$ is large), more sensors can be shut off
to save energy, since the remaining sensors have very good
channels.

In our model we minimize the power sum $\sum_k{P_k}$, \ie, the
$L^1$-norm of the transmission power vector $\mathbf{P}=(P_1, P_2,
\ldots, P_K)$. If the channel gain and the variance of the
observation noise for each sensor are ergodically time-varying on
a block-by-block basis, minimizing the $L^1$-norm of $\mathbf{P}$
in each time block minimizes $E\{\sum_k{P_k}\}$ with $E\{ \}$ the
expectation operation. In other words, it maximizes the average
node lifetime, which is defined as\\
$\frac{1}{K}\sum_k\frac{E_0}{E\{P_k\}}$ with $E_0$ the battery
energy available to each sensor (we assume that $E_0$ is the same
for all the sensors). This can be proved by the fact that
$\frac{1}{K}\sum_k\frac{E_0}{E\{P_k\}}\ge\frac{E_0}{E\{\frac{1}{K}\sum_k{P_k}\}}$.
However, when the channel is static and the variance of the
observation noise is time-invariant, minimizing the $L^1$-norm may
lead some individual sensors to consume too much power and die out
quickly. In this case minimizing the $L^\infty$-norm, \ie,
minimizing the maximum of the individual power values, is the most
fair for all sensors, but the total power consumption can be high.
As in~\cite{Jinjun_Shuguang1}, we can make a compromise to
minimize the $L^2$-norm of $\mathbf{P}$. In this way, we can
penalize the large terms in the power vector while still keeping
the total power consumption reasonably low. For the $L^2$-norm
minimization, the problem formulation becomes
\vspace{-5pt}\begin{eqnarray}\label{final_prob_norm2} \min &&
\sum_{k=1}^{K}\frac{W^4\xi_{k}^4}{g_k^2}\left(\frac{r_k}{1-r_k\sigma_k^2}\right)^2\nr
\\ \mbox{s.~t.} &&
\sum_{k=1}^Kr_k\ge\frac{1}{D_0};\hspace{0.3cm}
0\le{r_{k}}<{\frac{1}{\sigma_k^2}},\hspace{0.3cm}\forall{k},
\end{eqnarray}
which we can solve using interior point methods~\cite{Boyd1}. In
the next section, we compare the power efficiency of the analog
and the digital approaches previously discussed, where we minimize
the $L^2$-norm of $\mathbf{P}$. We use the $L^2$-norm since it
simplifies the comparison, but the comparison can be made for any
power norms.

\vspace{-10pt}
\section{Analog vs. Digital}\vspace{-5pt}

In order to transmit the observed analog signal with frequency
range $[-B/2,B/2]$ to the fusion center, each sensor in the
digital system proposed in~\cite{Jinjun_Shuguang1} first samples
the signal at a sampling rate $B$, then quantizes each sample into
$b_k$ bits, and finally uses uncoded MQAM to transmit the $b_k$
bits with a symbol rate $B$ and constellation size $M=2^{b_k}$.
Therefore, the total transmission bandwidth in the passband is
approximately equal to $KB$ for the digital system where TDMA is
used for the multiple access. For the SSB scheme used in our
analog system, each sensor only occupies $B/2$ in the passband
such that the total bandwidth requirement is $KB/2$ when FDMA is
used for the multiple access. Therefore, under the assumption of
orthogonal channel usage, the analog system can support double the
number of sensors that the digital system supports in the same
amount of bandwidth. This is mainly caused by the fact that the
digital approach proposed in~\cite{Jinjun_Shuguang1} forces the
transmission symbol rate to be equal to the sampling rate in the
source coding part.

\begin{figure}[!h]
\begin{center}
 {\includegraphics[width=2.8in, height=1.6in]{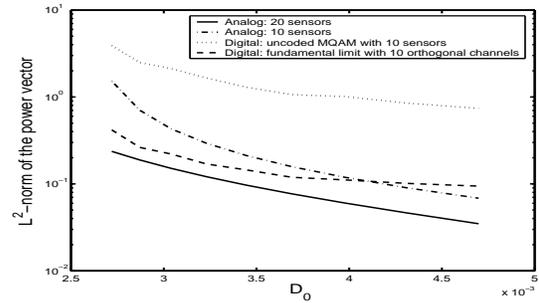}}
\end{center}
\vspace{-15pt} \caption{Comparison of the $L^2$-norm of the power
vector }\label{Fig_power_2norm_2N} \vspace{-15pt}
\end{figure}

The power efficiency comparison between the analog approach and
the digital approach is shown in Fig.~\ref{Fig_power_2norm_2N},
where we deploy $10$ sensors for the digital system and we plot
the power curves for both the case where uncoded MQAM is used (the
dotted line) and the case where single-user capacity-achieving
channel codes are applied to each transmitter, which gives the
fundamental Shannon limit (the dashed line) for the digital system
with orthogonal channel usage. From the figure we see that the
analog approach with $10$ nodes, which have the same observation
quality and transmission distances as in the digital system, is
more energy efficient than the digital approach with uncoded MQAM,
but not necessarily more energy efficient than the fundamental
limit curve. However, since we can support $20$ nodes in the
analog system with the same bandwidth requirement, with the $20$
analog nodes we can achieve a better power efficiency (the solid
line) than the digital system with $10$ nodes under the optimal
single-user channel coding. This is true even when the extra ten
nodes in the analog system have worse observation quality and
larger transmission distances than the first ten nodes.


\vspace{-15pt}
\section{Conclusions}\vspace{-5pt}
In this paper, we have shown that for a bounded source with
unknown statistics and a fusion center equipped with the best
linear unbiased estimator, we can minimize the total power
consumption across all the sensor nodes under a certain distortion
requirement. The information collection can be implemented with an
analog approach, which may be more energy efficient than the
digital approach when only orthogonal multiple access schemes such
as TDMA and FDMA are used.

\end{document}